\documentclass[twocolumn,aps,prc,reprint]{revtex4-1}
\usepackage{blindtext}
\usepackage{graphicx}
\usepackage{lipsum}
\usepackage{dcolumn}
\usepackage{bm}
\usepackage{booktabs}
\usepackage{color}
\usepackage{mathrsfs}
\usepackage{amsmath}
\usepackage{ulem}
\usepackage{verbatim}
\usepackage{indentfirst}
\usepackage{float}
\usepackage[colorlinks,
           bookmarks=true,
           linkcolor=blue,
           urlcolor=blue,
            anchorcolor=black,
            citecolor=blue
            ]{hyperref}
\usepackage[bf,raggedright,indentafter]{titlesec}
\usepackage{hypcap}
\usepackage{epsfig}

\begin{document}
\title{Data-driven subtraction of anisotropic flows in jet-like correlation studies in heavy-ion collisions}

\author{Liang Zhang$^{1,3}$}
\author{Kun Jiang$^{2,3}$}
\author{Cheng Li$^2$}
\author{Feng Liu$^1$}
\email[Electronic address:~]{fliu@mail.ccnu.edu.cn}
\author{Fuqiang Wang$^{3,4}$}
\email[Electronic address:~]{fqwang@purdue.edu}
\affiliation{%
$^1$ Key Laboratory of Quark and Lepton Physics (MOE) and Institute of Particle Physics,\\
Central China Normal University, Wuhan 430079, China\\
$^2$Department of Modern Physics, University of Science and Technology of China, Hefei 230026, China\\
$^3$Department of Physics and Astronomy, Purdue University, West Lafayette, Indiana 47907, USA\\
$^4$School of Science, Huzhou University, Huzhou, Zhejiang 313000, China
}

\begin{abstract}
\begin{description}
\item[Background] Measurements of two-particle azimuthal angle correlations are a useful tool to study the distribution of jet energy loss, however, they are complicated because of the significant anisotropic flow background.
\item[Purpose]We devise a data-driven method for subtracting anisotropic flow background in jet-like correlation analysis.
\item[Method] We first require a large recoil momentum ($P_x$) within a given pseudo-rapidity ($\eta$) range from a high-transverse momentum particle to enhance in-acceptance population of away-side jet-like correlations. Then we take the difference of two-particle correlations in the close-region and far-region with respect to the $\eta$ region of $P_x$ to subtract the anisotropic flow background.
\item[Results] We use a toy model which contains only anisotropic flow and PYTHIA8 which have jets to demonstrate the validity of our data-driven method.
\item[Conclusions] The results indicate that the data-driven method can subtract anisotropic flow effectively.

\end{description}
\end{abstract}
\maketitle

\section{Background}
Quark-gluon plasma (QGP) is believed to have been produced in relativistic heavy-ion collisions~\cite{Adams:2005dq,adcox2005formation,arsene2005quark,back2005phobos,Muller:2012zq}. An important evidence for the discovery of the QGP is jet quenching, which refers to the suppression of high transverse momentum ($p_{T}$) hadron yields and correlation amplitudes in heavy-ion collisions compared to those properly normalized in proton-proton collisions~\cite{Wang:1991xy}.
High-$p_T$ hadrons and jets are produced by hard (high transverse momentum transfer) processes which can be calculated by perturbative quantum chromodynamics (QCD). 
In vacuum, without nuclear effects like those found in the environment of heavy ion collisions, the cross sections of these processes are proportional to the number of nucleon-nucleon binary collisions~\cite{Jacobs:2004qv}. 
Deviation from this vacuum expectation, namely the suppression of high-$p_T$ hadrons, has been systematically studied over a wide range of collision energy and system size~\cite{Adams:2005dq,adcox2005formation,arsene2005quark,back2005phobos,Muller:2012zq}. Such studies have been extended to high $p_T$ jets~\cite{Muller:2012zq}.

Jets are produced in pairs by hard processes in leading order QCD. Two-particle jet-like angular correlations are a useful tool to study interactions between jets and the QGP medium, the underlying physics mechanism for jet quenching~\cite{Jacobs:2004qv}.
While many insights have been gained through high-$p_T$ correlations, the information from those studies is limited because jet energy loss is mostly populated at low $p_T$~\cite{Jacobs:2004qv, Wang:2013qca}.
However, at low $p_{T}$ where jet-quenching effects are the strongest, two-particle correlations are contaminated by large anisotropic flow backgrounds~\cite{Wang:2013qca}. Subtraction of flow backgrounds suffers from large uncertainties in the anisotropy measurements together with the large combinatorial background level.

\section{Purpose and Method}
\begin{figure}[H]
   \centering
    \includegraphics[width=0.45\textwidth]{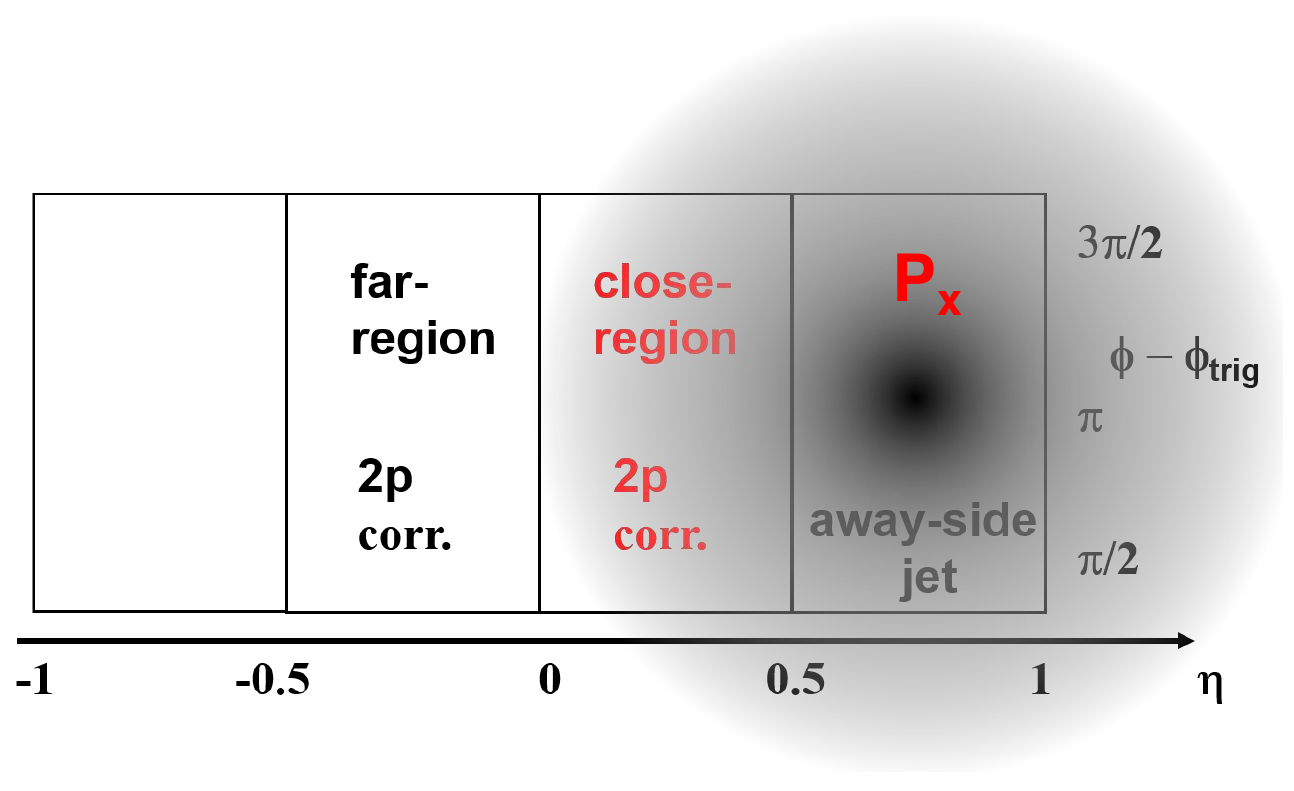} 
   \caption{Cartoon for close region and far region of the data-driven method. The requirement of a large recoil $P_x$ (Eq. \ref{eq:Px}) in a specific $\eta$ region ($0.5 < \eta < 1$ shown here) selects events to enhance away-side jet population in the acceptance (indicated by the dark blob).}
   \label{fig:deta_dphi}
\end{figure}

In order to circumvent the large flow background problem, we devise a new method to subtract the flow backgrounds using data themselves. The method is as follows. We enhance away-side jet population in acceptance by requiring a large recoil transverse momentum ($P_{x}$) to a high $p_T$ trigger particle (3 $<p_T^{trig}<$ 10 GeV/c). $P_{x}$ is the sum of projections, along the trigger particle direction, of the $p_T$'s of all charged particles in the opposite azimuthal hemisphere of the trigger within a specific pseudo rapidity ($\eta$) range, namely
\begin{eqnarray}\label{eq:Px}
&P_{x}|^{\eta_2}_{\eta_1}= &\sum_{\eta_1 < \eta < \eta_2, |\phi-\phi^{trig}| > \pi/2} p_T \cos(\phi-\phi^{trig}).
\end{eqnarray}
Here $\phi^{trig}$ and $\phi$ are the azimuthal angles of the trigger particle and the charged particle in the opposite hemisphere. The $p_T$ range of the charged particles is wide open to include essentially all particles in the $P_x$ calculation.
Practically, for each centrality bin and a given $\eta_1<\eta<\eta_2$ region, we obtain the $P_x$ distribution and select a small fraction (e.g. 10\%) of events with the highest $|P_{x}|$. The selected events should have a relatively high probability to have a significant away-side jet contribution inside the detector acceptance. Figure~\ref{fig:deta_dphi} illustrates the method by depicting the away-side acceptance in $\eta$ and $\phi$. The $\eta$ range for the $P_x$ calculation is $0.5<\eta<1$ in this illustration.

We define two regions in $\eta$ symmetric about the mid-rapidity, one called ``close region'' that is close to the $P_x$ region (the $\eta$ region where the $P_x$ is calculated) and the other called ``far region'' that is far from the $P_x$ region. Literately, the ``close" and ``far'' here are used to describe the distance between a specific $\eta$ region and the $\eta$ region where the away-side jet population is enhanced. 
We analyze two-particle correlations between trigger particles and associated particles in the close and far regions separately. 
The anisotropic flow contributions are nearly equal because the two regions are symmetric about the mid-rapidity.
They should cancel in the difference of the correlations in close region and far region.
The away-side jets, on the other hand, have a more significant contribution to close-region than to far-region due to the different $\eta$ distances. 
The difference of close- and far-region two-particle correlation contains only the contribution of away-side jet-like correlations. \\

This method has been applied to STAR data analysis~\cite{KunProceedings, LiangProceedings}.
Two $\eta$ regions $-1 < \eta < 0.5$ and $0.5 < \eta <1$ are used for $P_x$ calculation. For each of these, the close and far regions are properly defined. Events that satisfy the 10\% cut in both $P_x|_{-1}^{-0.5}$ and $P_x|_{0.5}^{1}$ are discarded because the correlation difference between close and far regions would be canceled. It is worthwhile to note that, due to flow fluctuations and the selection of the large $P_x$ events, the flow contributions to two-particle correlations in the close region and the far region can differ slightly. This effect can be accessed as a systematic uncertainty by varying the $\eta$ regions as was done in the STAR analysis~\cite{KunProceedings, LiangProceedings}.

\section{Results}
In this paper we use two models to demonstrate the feasibility of our method. One is a toy model simulation where only elliptic flow is included but no jets. The other is the PYTHIA model, where jets are present but no anisotropic flows.

\subsection{Toy model with flow only}
\label{Toy model with flow only}
We construct a toy model which contains only the $2^{nd}$-order anisotropic flow. The kinematic distribution of particles is given by
\begin{eqnarray}
\frac{d^{3}N}{p_{T}d\eta dp_{T}d\phi} &\propto& e^{-\frac{p_{T}}{T}}(1+2v_{2}\cos2\phi),
\end{eqnarray}
where we set T = 0.3 GeV/c and $v_{2}$ = 0.05. We generate a sample of 100 million events and each event has 1000 particles.

\begin{figure}[H]
   \centering
  \includegraphics[height=6cm]{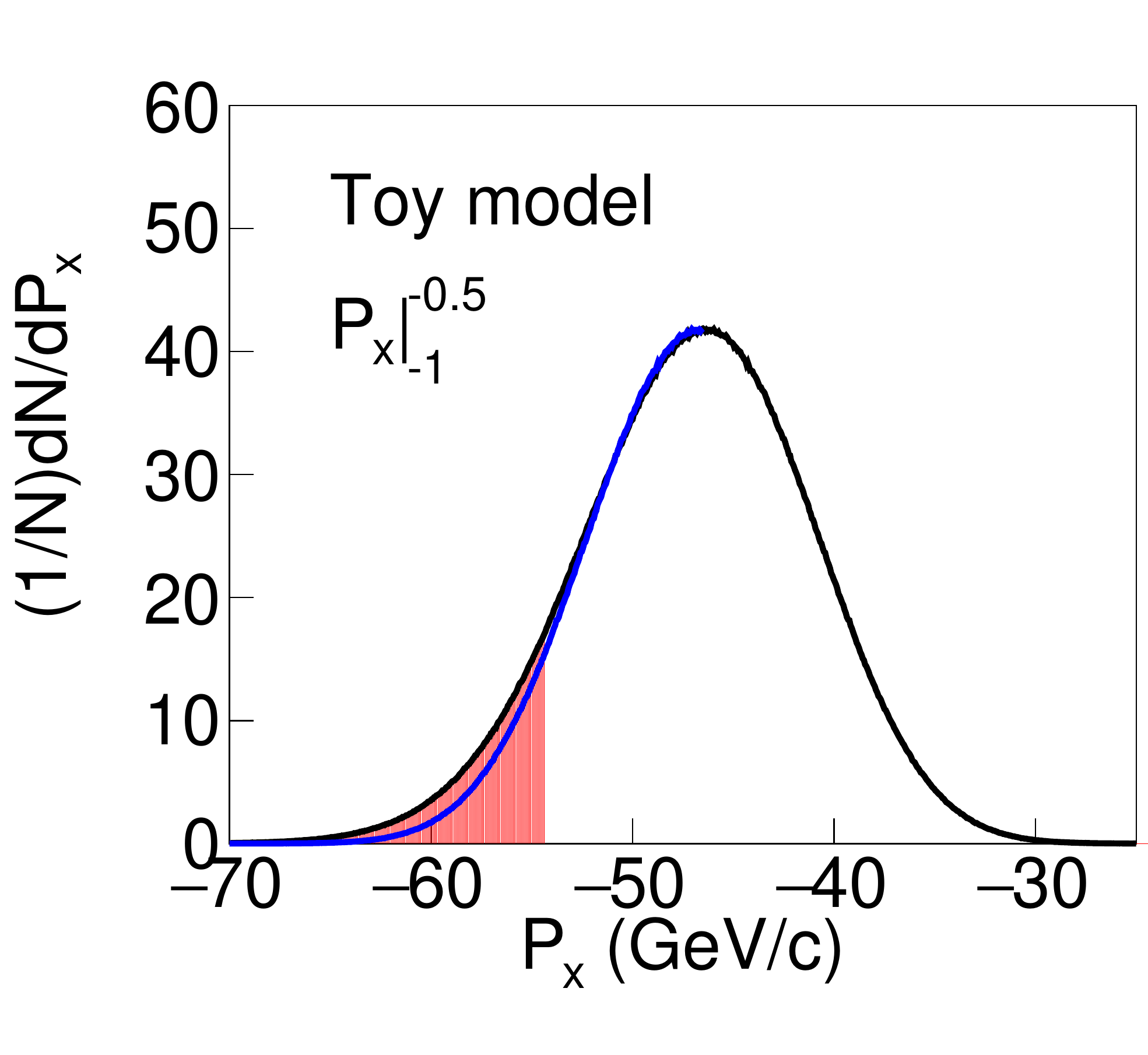} %
     \caption{The recoil $p_{T}$ distribution from the toy model simulation for high $p_T$ trigger particles of 3$~<p_{T}^{trig}<~$10 GeV/c. The red shaded area is the 10\% largest $|-P_x|$ events. The right side of the distribution to the maximum is reflected as the blue curve.}
   \label{fig:Px_toy}
\end{figure}

We select the $\eta$ range within $(-1, -0.5)$ and $(0.5, 1)$ for $P_{x}$ calculation. Figure~\ref{fig:Px_toy} shows the distribution of $P_{x}|_{-1}^{-0.5}$ as an example. 
The $P_x$ distribution is asymmetric as indicated by the blue curve in Fig.~\ref{fig:Px_toy}.
The asymmetry is simply a feature of the $P_x$ distribution.

\begin{figure*}[htbp]
   \centering
  \includegraphics[width=\textwidth]{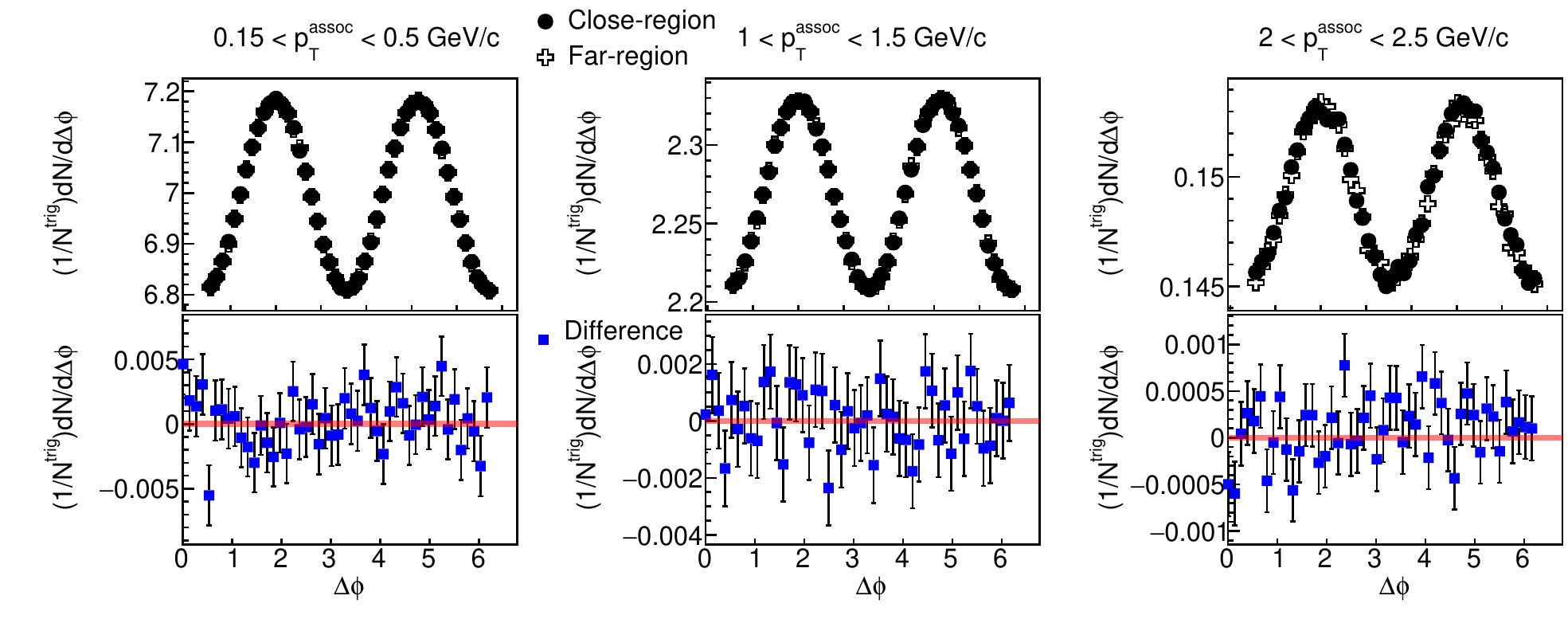} 
    \caption{Upper panels: Close-region and far-region two-particle correlations relative to the trigger particles (3$~<p_{T}^{trig}<~$10 GeV/c) for different $p_{T}^{assoc}$ bins from the toy model. Lower panels: The corresponding difference between close- and far-region two-particle correlations.}
   \label{fig:toy_correlations}
\end{figure*}

The upper panels of Fig. \ref{fig:toy_correlations} show two-particle azimuthal angle correlations for close- and far-region in three selected $p_{T}^{assoc}$ bins from the toy model simulation; $\Delta\phi$ is the azimuthal angle difference between the associated and trigger particles. Since the toy model contains only elliptic flow without any jets, there should be no jet-like correlation signals. Moreover, even though a large $P_x$ is selected (out of statistical fluctuations), the close and far regions are independent of the $P_x$ region and are not affected. Therefore, the close- and far-region two-particle correlations should be equal. Indeed they are found to be the same. The lower panels of Fig. \ref{fig:toy_correlations} show the difference between close- and far-region two-particle correlations. The difference is consistent with zero within statistical errors.
\subsection{PYTHIA8}
We employ PYTHIA8 (version8.235) as our jet model to illustrate the validity of our data-driven method. PYTHIA is an event generator to simulate relativistic collision events between elementary particles like $e^{\pm}$, p and $\bar{p}$. A key improvement of PYTHIA8 from earlier versions is the multi-partonic interaction mechanism. Multi-partonic interactions are essiential to explain the underlying event, the multiplicity distribution and the flow-like patterns in elementary particle collisions. The details of the physics processes in PYTHIA8 can be found in Ref.~\cite{Sjostrand:2007gs}. 

\begin{figure}[H]
   \centering
  \includegraphics[width=0.35\textwidth]{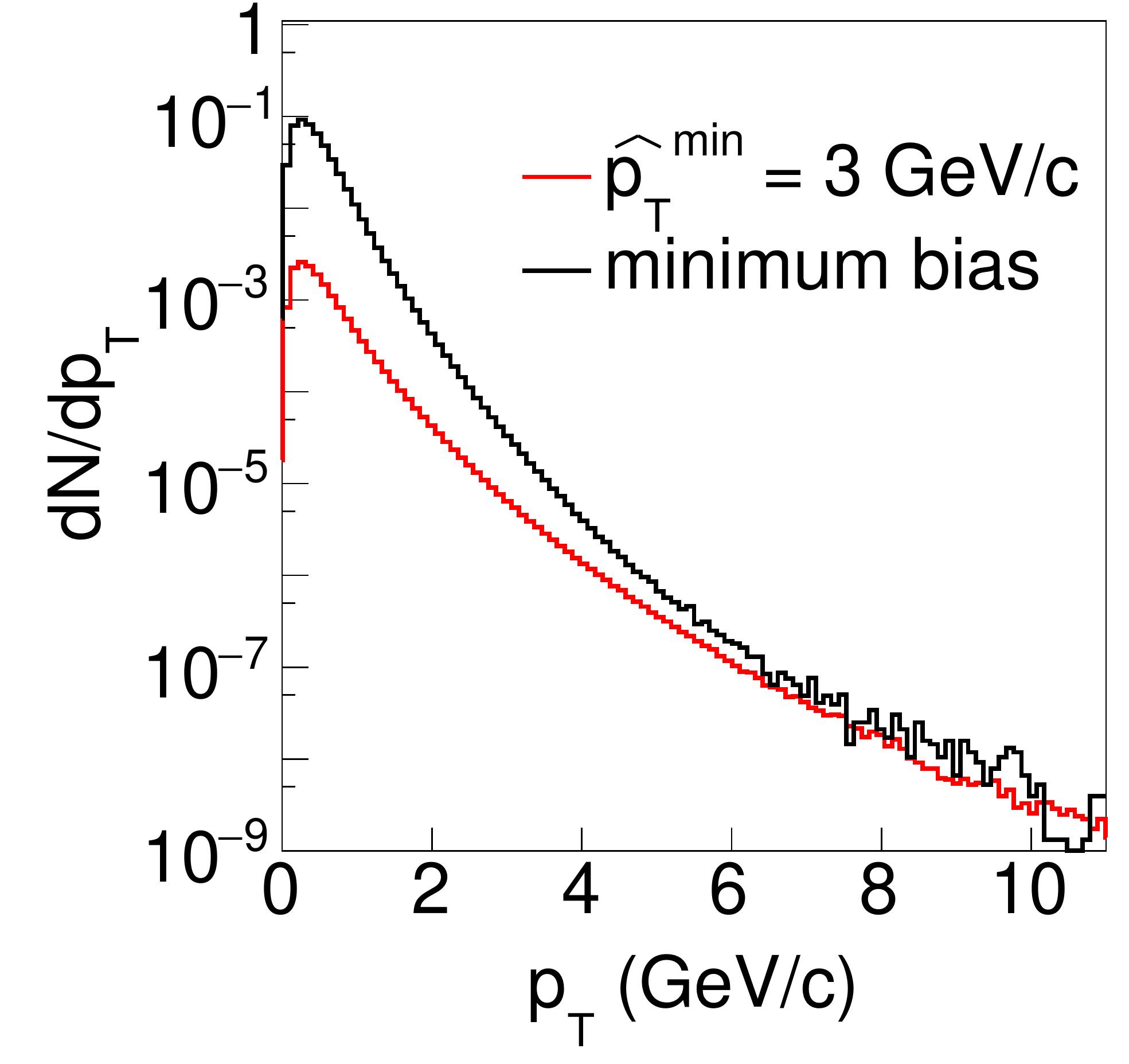} 
   \caption{The $p_{T}$ distributions of charged particles at mid-rapidity ($|\eta|<$ 1) from minimum bias PYTHIA8 and from biased events with $\hat{p_{T}}^{min} = 3$~GeV/c. The distributions are scaled by their respective cross sections. }
   \label{fig:com_pT}
\end{figure}

We run proton-proton collisions at $\sqrt{s}$=200~GeV.
We turned on the inelastic, non-diffractive component of the total cross section for all hard QCD process (HardQCD: all = on). We set $\hat{p_{T}}^{min}=3$ GeV/c to enhance the population of high $p_{T}$ particles. For minimum biased PYTHIA8, $\hat{p_{T}}^{min}$ is 0 GeV/c.
We generate a billion p+p events at $\sqrt{s}=200$~GeV. 
Figure \ref{fig:com_pT} depicts the $p_{T}$ distributions from minimum bias (MB) events and biased events with $\hat{p_{T}}^{min}=$3 GeV/c. The distributions are normalized according to their respective cross sections: $
\sigma_{\text{pp}}$ = 42 mb for MB and $\sigma_{pp}$ = 1.9637 $\pm$ 0.0003 mb for 3 GeV/c $\hat{p_{T}}^{min}$.

\begin{figure}[H]
   \centering
    \includegraphics[width=0.35\textwidth]{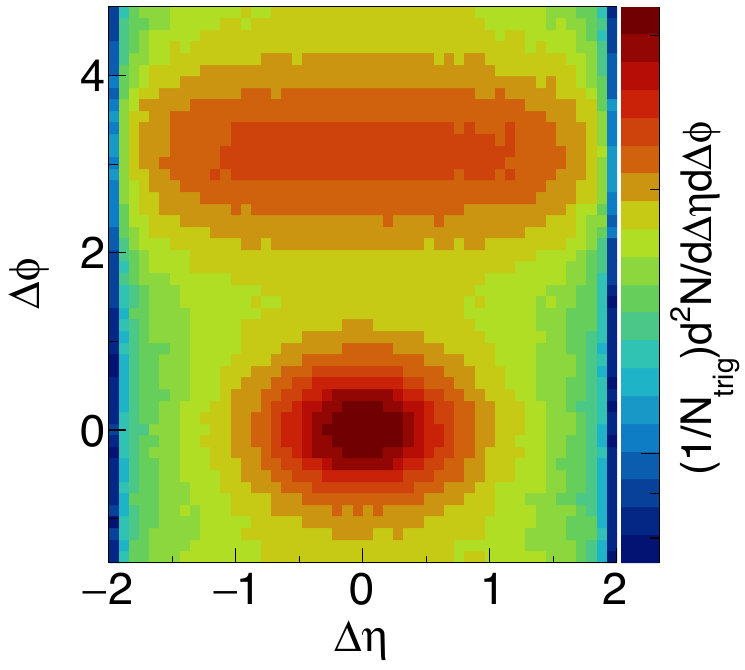} 

   \caption{Two-particle ($\Delta\phi,~\Delta\eta$) correlations with 3 $<p_{T}^{trig}<$ 10 GeV/c and 1 $<p_{T}^{assoc}<$ 1.5 GeV/c in p+p collisions at $\sqrt{s}$ = 200~GeV from PYTHIA8. The $\eta$ range of both the trigger and associated particles are $|\eta|<1$.}
   \label{fig:deta_dphi_2}
\end{figure}

Figure~\ref{fig:deta_dphi_2} shows the two-particle correlations in ($\Delta\phi,~\Delta\eta$) with 3 $<p_{T}^{trig}<$ 10 GeV/c and 1 $<p_{T}^{assoc}<$ 1.5 GeV/c. The triangular acceptance in $\Delta\eta$ has been corrected. The dominant feature here is the jet correlations. The near-side jet is concentrated at $(\Delta\eta,~\Delta\phi)=(0,~0)$ and the away-side jet at $\Delta\phi\approx\pi$ contributes approximately uniformly in $\Delta\eta$ within $\Delta\eta<1.5$.\\
\begin{figure*}[htbp]
   \centering
  \includegraphics[width=\textwidth]{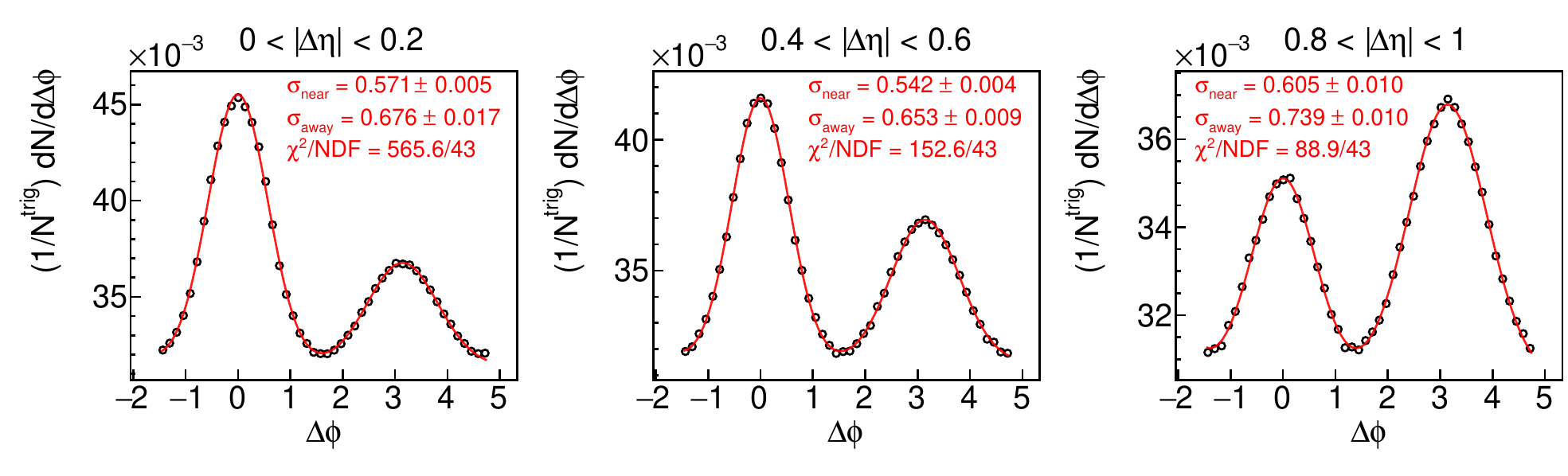} %
     \caption{Two-particle azimuthal correlations of selected $\Delta\eta$ bins in p+p collisions at $\sqrt{s}$ = 200~GeV from PYTHIA8 ($1<p_{T}^{assoc} <1.5$~GeV/c, $3<p_{T}^{trig} <10$~GeV/c). Red curves are "double Gaussian + constant" fits.}
   \label{fig:correlation_old}
\end{figure*}

In PYTHIA8 there is no anisotropic flow in the underlying event.
Jet-like correlations are relatively easy to access without the need of our sophisticated data-driven method.
An important assumption in our data-driven method is that the jet-like $\Delta\phi$ correlation width is independent of $\Delta\eta$.
We want to check the validity of this assumption by analyzing jet-like correlation in PYTHIA8 using the standard method.
Figure~\ref{fig:correlation_old} shows the simple two-particle azimuthal correlations in $3<p_{T}^{trig}<$ 10 GeV/c and 1$~<~p_{T}^{assoc}~<~$1.5 GeV/c. Different sub-panels give the correlations in different bins of $\eta$ gap between trigger and associated particles ($\Delta\eta$). The data points are fitted by the ``periodic double Gaussian + constant'' functions as shown by the solid red curves. The $\chi^{2}$/NDF of the fits are large, which implies that Gaussian may not be the best function to describe these peaks.
Our statistical uncertainties are small which make the deviation from Gaussian easily observable, resulting in a large $\chi^{2}$/NDF. Nevertheless, the periodic double Gaussian seems to describe near-side and away-side peaks reasonably well. Since we are focusing on the peaks widths instead of shapes, the periodic double Gaussian function is decent to give a faithful width representation.
\begin{figure}[!htb]
   \centering
  \includegraphics[height=6cm]{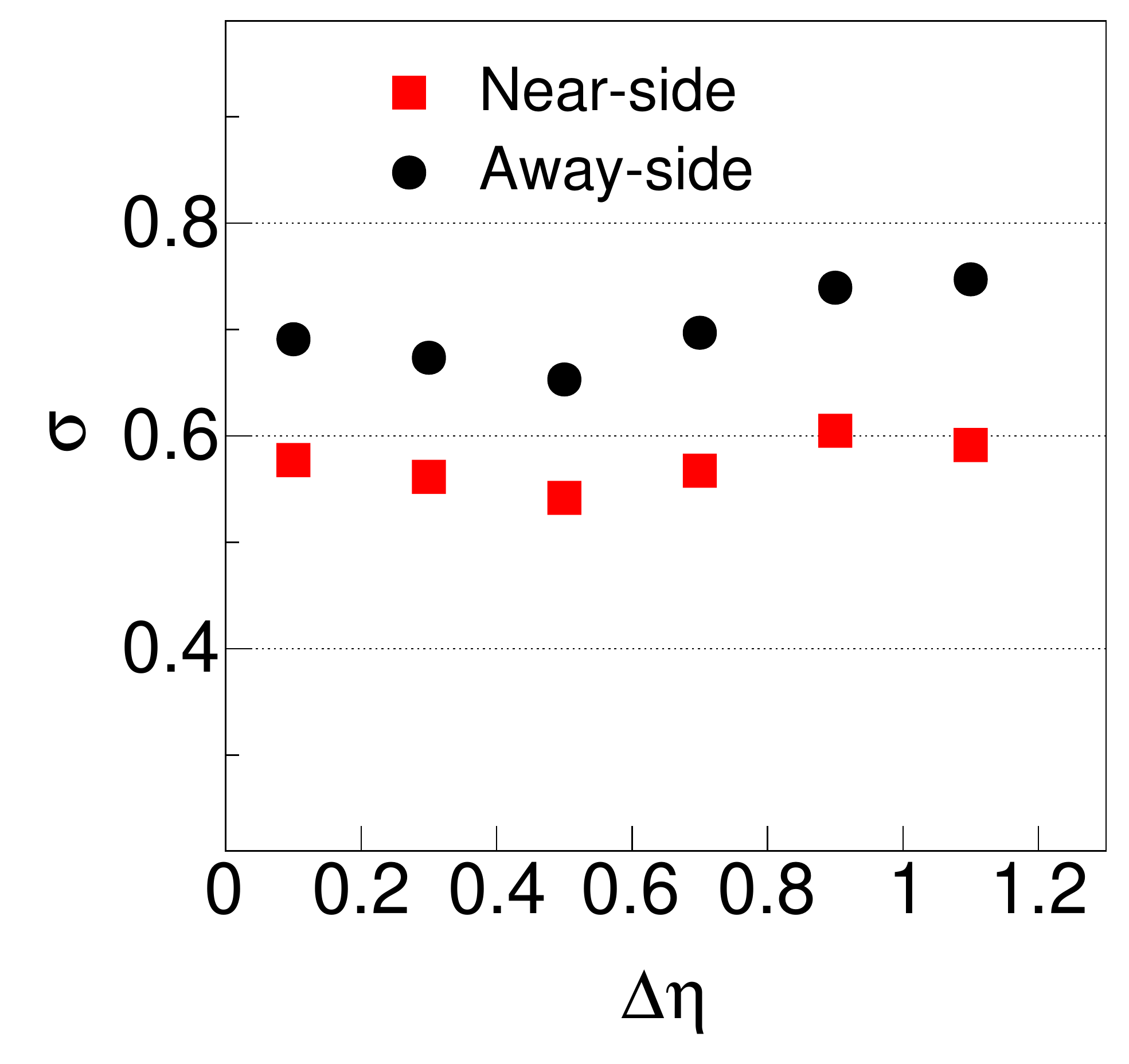} %

     \caption{Near- and away-side correlation widths ($\sigma$) as a function of $\Delta\eta$ from ``periodic double Gaussian + constant'' function fit. The trigger and associated particle $p_T$ ranges are $3<p_{T}^{trig} <10$~GeV/c and 1$~<~p_{T}^{assoc}~<~$1.5~GeV/c, respectively. 
     }
   \label{fig:sigma_vs_deltaeta}
\end{figure}

We study the near- and away-side correlations width as a function of $\Delta\eta$ as shown in Fig. \ref{fig:sigma_vs_deltaeta}.
Both the near- and away-side peak widths decrease with $\Delta\eta$ slightly at small $\Delta\eta$ and then increase with $\Delta\eta$ at larger $\Delta\eta$. This appears to be a feature of PYTHIA8, the physics reason of which is beyond the scope of investigation in this paper.
We focus on whether the jet $\Delta\phi$ width is constant over $\Delta\eta$ from the jet axis, hence the near-side width is most relevant. In our data-driven method, we enhance the jet population in the $\eta$ region of $P_x$, so the away-side jet axis is presumably somewhere close to the $P_x$ region.
The close-region $\Delta\phi$ correlation is then integrated over a range in $\Delta\eta$ from the jet axis.
The far-region $\Delta\phi$ correlation is also integrated over a $\Delta\eta$ range, but relatively further away.
In other words, the close-region $\Delta\phi$ width can be regarded as an average of the near-side peak width in Fig.~\ref{fig:sigma_vs_deltaeta} over a range of $\Delta\eta$ at the small $\Delta\eta$ side, and the far-region $\Delta\phi$ width can be regarded as an average over a range at the large $\Delta\eta$ side.
Figure~\ref{fig:sigma_vs_deltaeta} shows that the width of near-side jet-like correlations in $\Delta\phi$ is nearly constant over $\Delta\eta$, which indicates that our assumption of the same close-region and far-region correlation widths is reasonable.
\begin{figure}[!htb]
   \centering
  \includegraphics[height=6cm]{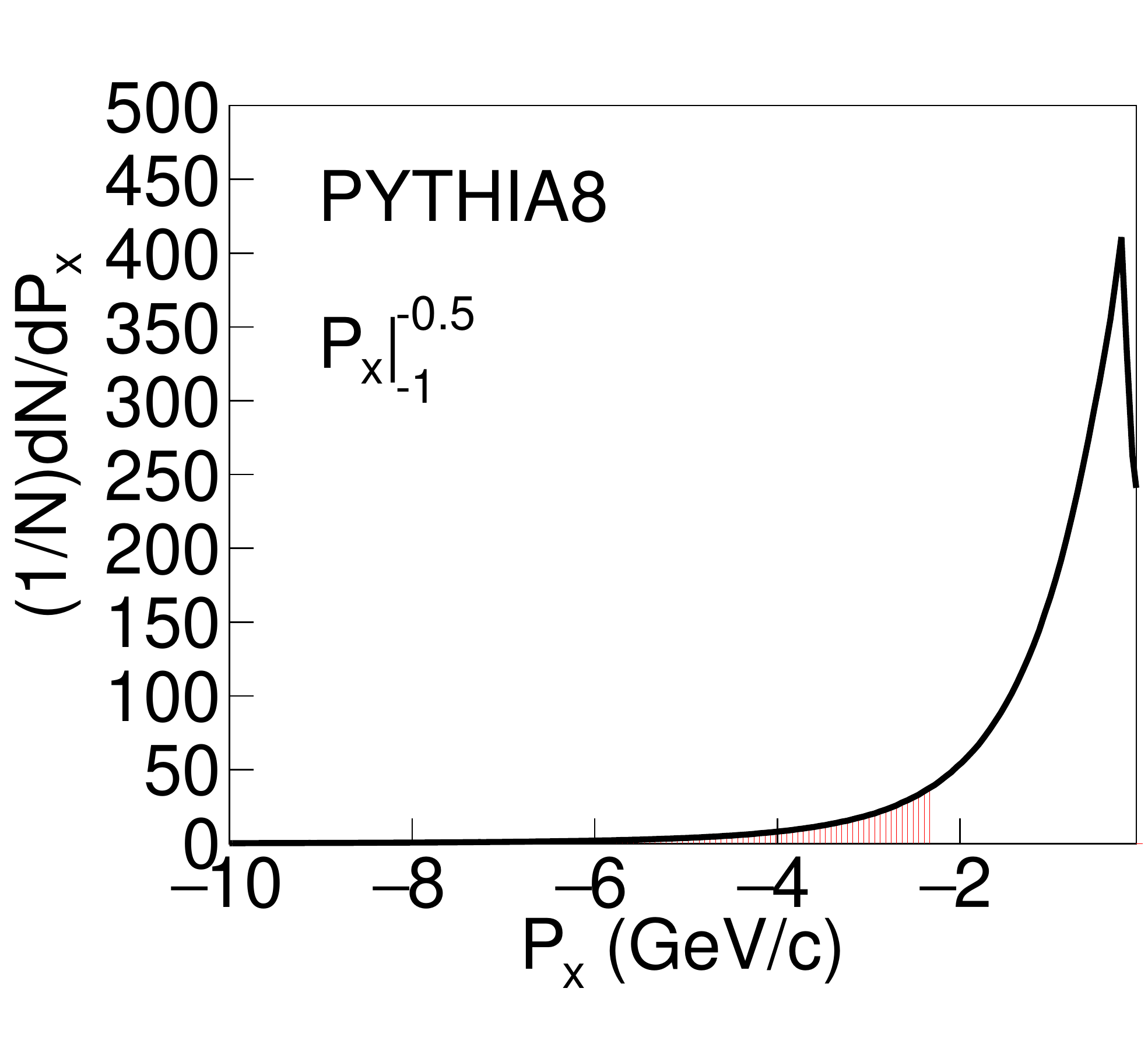} %
     \caption{Same as Fig.~\ref{fig:Px_toy} but from PYTHIA8.}
   \label{fig:Px_pythia}
\end{figure}
Now we turn to our data-driven method to analyze the PYTHIA8 data.
Figure~\ref{fig:Px_pythia} shows the distribution of $P_{x}|_{-1}^{-0.5}$ from PYTHIA8. The asymmetry from PYTHIA8 is significant because of the small multiplicity. The long negative $P_x$ tail is predominately from away-side jet contributions. We select the 10\% of total events with the highest $-P_{x}$ to enhance away-side jet population in the corresponding $\eta$ regions.

\begin{figure*}[htbp]
   \centering
  \includegraphics[width=\textwidth]{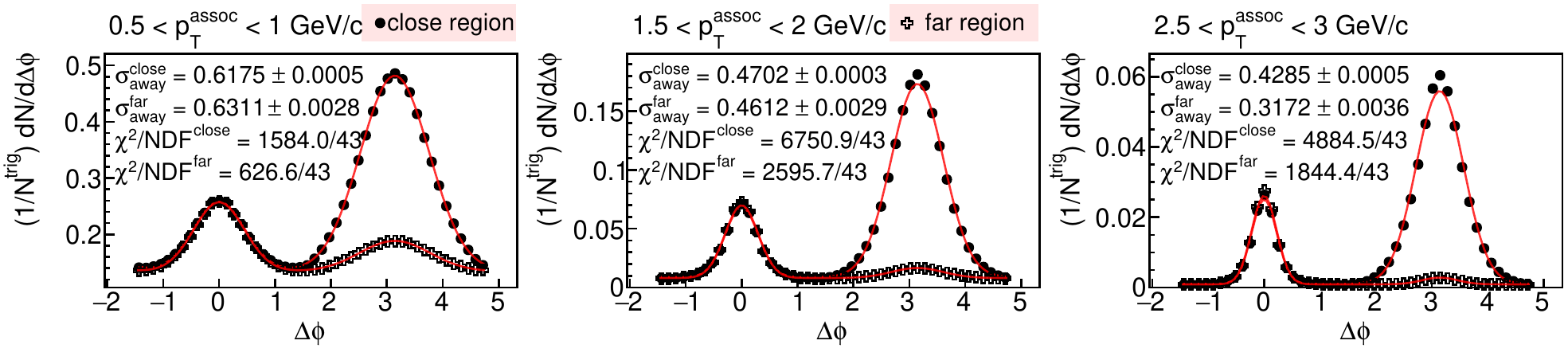} 
   \caption{Close-region and far-region two-particle correlations from PYTHIA8. The trigger $p_T$ is 3$~<~p_{T}^{trig}~<~$10 GeV/c. The red curves are ``periodic double Gaussian + constant'' fits.}
   \label{fig:pythia_correlations_dg}
\end{figure*}

We calculate the close- and far-region two-particle correlations with respect to the trigger particles (3$~<p_{T}^{trig}<~$10 GeV/c) for three selected $p_{T}^{assoc}$ bins in Fig.~\ref{fig:pythia_correlations_dg}. The near-side correlations are equal between close region and far region. The away-side correlation in the close region is larger than the far region because of the larger jet-like contributions to the close region. The results are fitted by the ``periodic double Gaussian + constant'' function as the red curves shown. The $\chi^{2}$ of the fits are large. The correlations are not well described by a double Gaussian function, especially the away-side peak at high-$p_T$. The fit to the far-region correlation is generally better than that to the close-region correlation. We also try to fit the away-side jet-like correlation with a single Gaussian function in narrow $\Delta\phi$ ranges around the peak, and found the close- and far-region away-side correlations are similar in width.

\begin{figure*}[htbp]
   \centering
  \includegraphics[width=\textwidth]{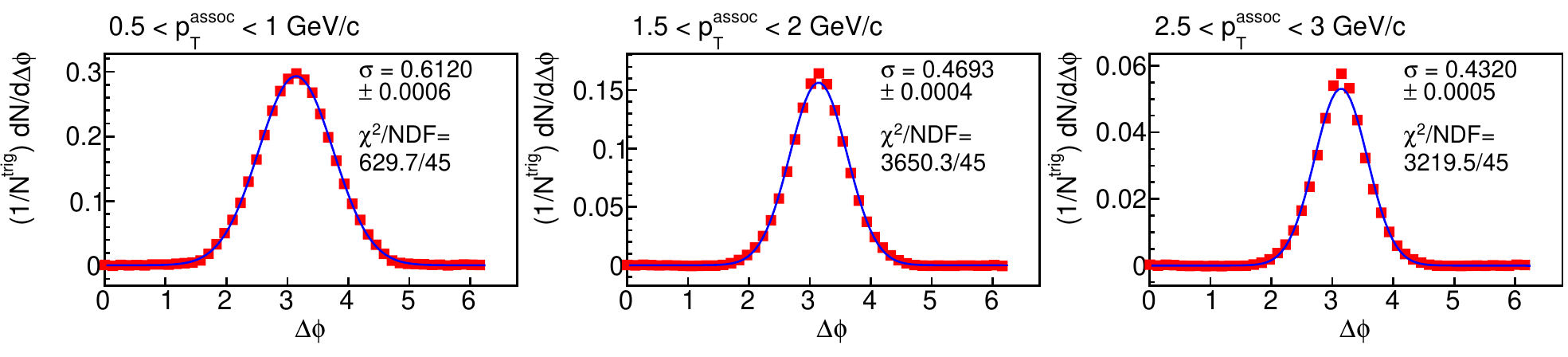} 
  \caption{The differences of two-particle correlations between close- and far-region from PYTHIA8. The trigger $p_T$ is 3$~<~p_{T}^{trig}~<~$10 GeV/c. The blue curves are ``Gaussian + constant" fits. }
  \label{fig:pythia_differences}
\end{figure*}

Figure~\ref{fig:pythia_differences} shows the differences of close- and far-region two-particle correlations in different $p_{T}^{assoc}$ bins. The blue curves are ``Gaussian + constant'' fits.
Again the fit $\chi^{2}$/NDF is large. The Gaussian does not seem to describe the correlation difference well, similarly to the fit description in Fig.~\ref{fig:pythia_correlations_dg}. This is especially apparent for the high $p_T$ bins, where the Gaussian fit curves fall below the PYTHIA8 data points.

\begin{figure}[!htb]
   \centering
 \includegraphics[height=6cm]{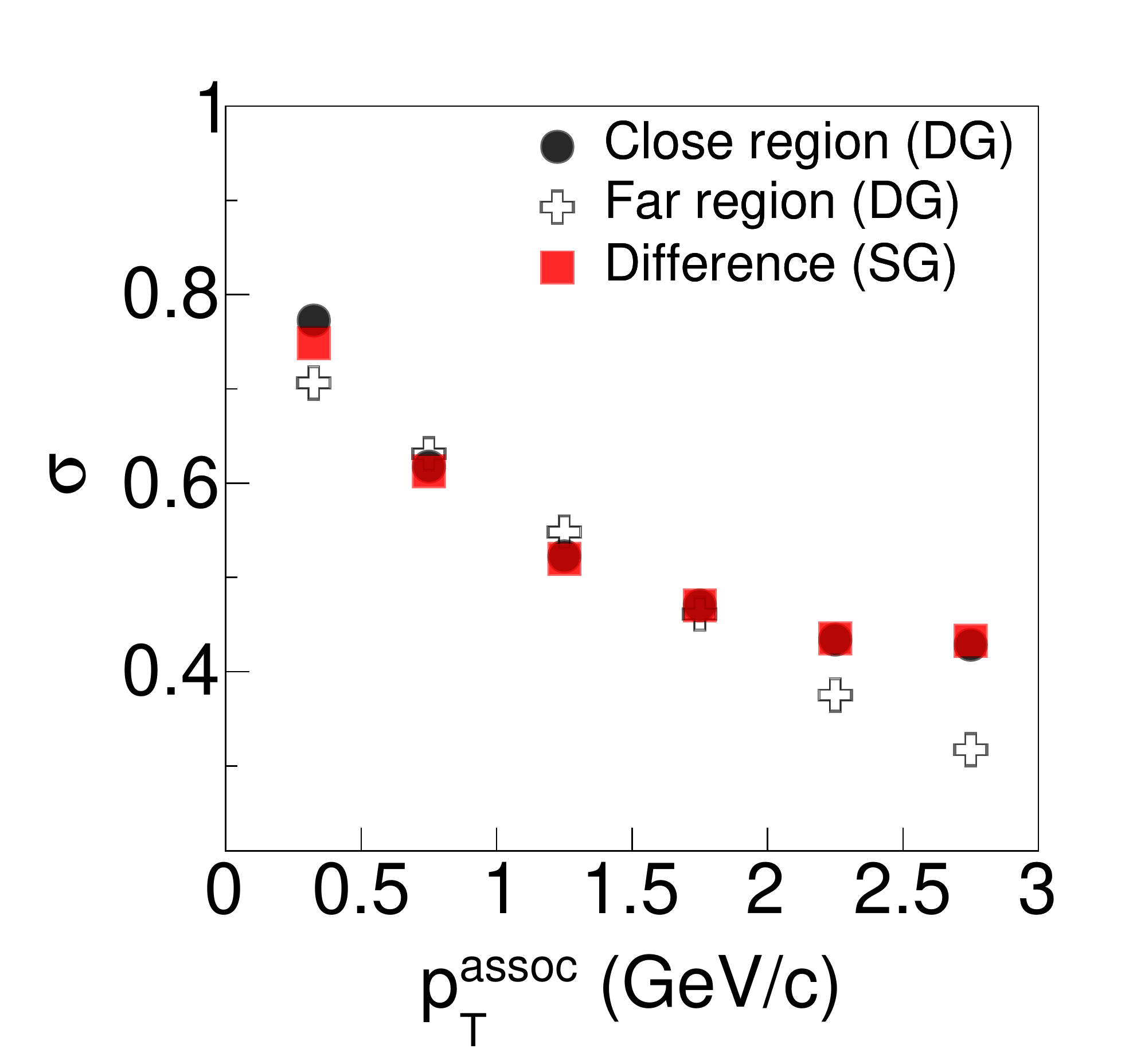} 
    \caption{Away-side correlation width ($\sigma$) as a function of $p_{T}^{assoc}$ from PYTHIA8 (Single Gaussian). 
    }
   \label{fig:pythia_sigma_vs_pT_narrow}
\end{figure}

Fig.~\ref{fig:pythia_sigma_vs_pT_narrow} shows the comparison of the Gaussian widths ($\sigma$) between close- and far-region correlations and their difference as function of $p_{T}^{assoc}$.
The correlation widths decrease with increasing $p_{T}^{assoc}$, as expected from hard processes.
The far-region correlation widths appear to be smaller than the close-region correlation widths at high $p_T$. This is because the Gaussian fit to the far-region correlation is better at the peak, than the close-region fit. This contributes to the better $\chi^{2}$ of far-region correlation fit. The good fit is probably because the correlation amplitude is small, so the Gaussian seems able to capture both the peak region and the rest tail regions. If we fit a Gaussian to a narrow range in $\Delta\phi$ around the peak, then the fit widths are equal between close-region and far-region. 

The widths of the close-region correlation and the correlation difference are well consistent in all $p_{T}^{assoc}$ bins. This is because the far-region correlation amplitude is relatively small, so the difference is dominated by the close-region correlation. At high $p_T$, the width of far-region jet-like correlation is not very important. However, the far-region jet-like correlation is still crucial for subtracting flow background in heavy-ion collisions.

\section{Conclusions} 
We have devised a data-driven method to subtract anisotropic flow background in jet-like correlation analysis. The method applies a lower cut on the recoil $P_x$ from a high-$p_T$ trigger particle to enhance away-side population inside acceptance. The jet-like correlation functions are constructed in two regions symmetric about mid-rapidity but with different $\eta$ distances from the $\eta$ region where the $P_x$ is calculated. By taking the difference, the anisotropic flow is mostly canceled and the remaining signal reflects the away-side jet-like correlation shape.
We have used a toy model which contains only $2^{nd}$ harmonic anisotropic flow, and PYTHIA8 which has (almost) only jets, to demonstrate the validity and feasibility of the data-driven method. With the toy model, we found the differences of close- and far-region away-side correlations in all $p_T^{assoc}$ bins to be well consistent with zero. This indicates that the data-driven method can subtract flow background effectively. 
With PYTHIA8, the results show a significant amplitude differences between close- and far-region two-particle correlations on the away-side, which indicates jet contributions. The differences of close- and far-region two-particle correlations are fitted by Gaussian functions. The Gaussian widths are found to faithfully represent the jet-like correlation peaks.
These results indicate that our data-driven method can subtract the flow background effectively and retain the jet-like correlation shape on the away side.

\section{Acknowledgments}
We thank Terrence Edmonds for proof reading the manuscript. This work is supported by the MoST of China 973-Project No.~2015CB856901, the National Natural Science Foundation of China under Grants No.~11890711, No.~11705194, No.~11847315, the Anhui Provincial Natural Science Foundation under Grant No.~1808085QA23, and the U.S.~Department of Energy (Grant No.~de-sc0012910).


\end{document}